\documentclass[aps,prl,twocolumn,longbibliography,superscriptaddress,floatfix,10pt]{revtex4-1}

\usepackage[normalem]{ulem}
\usepackage{amsfonts,amssymb,dsfont}
\usepackage[sumlimits,intlimits]{amsmath}
\usepackage{graphics}
\usepackage{graphicx}
\usepackage[dvipsnames]{xcolor}
\usepackage{color}
\usepackage{mathrsfs}
\usepackage{textcomp}
\usepackage{verbatim}
\usepackage[colorlinks=true,citecolor=blue]{hyperref}
\usepackage{bm}
\usepackage{soul}
\usepackage{braket}
\usepackage{times}
\usepackage[T1]{fontenc} 
\usepackage{booktabs}
\usepackage{lipsum}

\usepackage[final]{pdfpages}
\usepackage{pgffor}
\makeatletter
\AtBeginDocument{\let\LS@rot\@undefined}
\makeatother

\usepackage[capitalise,compress]{cleveref}
\crefrangelabelformat{equation}{\textup{(#3#1#4)}--\textup{(#5#2#6)}}

\DeclareMathOperator{\Tr}{{\rm Tr}}

\begin{document}

\title{Optimal control for quantum detectors}

\author{Paraj Titum}
\affiliation{Johns Hopkins University Applied Physics Laboratory, Laurel, Maryland 20723, USA}
\affiliation{Joint Quantum Institute, NIST/University of Maryland, College Park, MD 20742}

\author{Kevin M. Schultz}
\affiliation{Johns Hopkins University Applied Physics Laboratory, Laurel, Maryland 20723, USA}

\author{Alireza Seif}
\affiliation{Joint Quantum Institute, NIST/University of Maryland, College Park, MD 20742}
\affiliation{Joint Center for Quantum Information and Computer Science, NIST/University of Maryland, College Park, MD 20742}
\affiliation{Department of Physics, University of Maryland, College Park, MD 20742}

\author{Gregory D. Quiroz}
\affiliation{Johns Hopkins University Applied Physics Laboratory, Laurel, Maryland 20723, USA}

\author{B. D. Clader}
\affiliation{Johns Hopkins University Applied Physics Laboratory, Laurel, Maryland 20723, USA}


\begin{abstract}
Quantum systems are promising candidates for sensing of weak signals as they can provide unrivaled performance when estimating parameters of external fields. However, when trying to detect weak signals that are hidden by background noise, the signal-to-noise-ratio is a more relevant metric than raw sensitivity. We identify, under modest assumptions about the statistical properties of the signal and noise, the optimal quantum control to detect an external signal in the presence of background noise using a quantum sensor. Interestingly, for white background noise, the optimal solution is the simple and well-known spin-locking control scheme. We further generalize, using numerical techniques, these results to the background noise being a correlated Lorentzian spectrum. We show that for increasing correlation time, pulse based sequences such as CPMG are also close to the optimal control for detecting the signal, with the crossover dependent on the signal frequency. These results show that an optimal detection scheme can be easily implemented in near-term quantum sensors without the need for complicated pulse shaping. 
\end{abstract}

\maketitle
{\it Introduction.}--- Quantum systems are extremely sensitive to the environment which makes them an ideal candidate as a sensor of weak external fields. Many promising candidates have been put forward as quantum sensors such as defect centers in diamond or Silicon Carbide, SQUID based sensors, atomic sensors, along with many others (see e.g. \cite{RevModPhys.89.035002}). A variety of sensing techniques have been developed which can be used to detect either the magnitude or the phase of the signal. In Ramsey interferometry \cite{PhysRev.78.695}, a qubit is prepared in the equal superposition state and its oscillation frequency is sensitive to the splitting of the qubit, which depends on the external field to be sensed. This allows for estimation of the magnetic field amplitude with sensitivity limited by the free-evolution dephasing time of the qubit, which can be enhanced through optimal control methods \cite{PhysRevX.8.021059}. Similarly, detecting AC signals is possible with Ramsey and echo sequences such as Carr-Purcell (CP) \cite{PhysRev.94.630} and Dynamical Decoupling (DD) \cite{PhysRevA.58.2733, PhysRevLett.82.2417} sequences. These sequences can also be used for estimating frequencies of the signal \cite{Fedder2011, Aiello2013}. For amplitude sensing, the typical figure of merit one considers when measuring the performance of a quantum sensor is the sensitivity, which characterizes the smallest external field that can be measured in a given amount of time~\cite{RevModPhys.89.035002}. This can be formulated in terms of the quantum Cram\'er-Rao bound and has associated applications and limiting cases \cite{BRAUNSTEIN1996135, PhysRevLett.72.3439, doi:10.1142/S0219749909004839, PhysRevLett.106.090401, PhysRevLett.96.010401}.
\begin{figure}
\includegraphics[width=\linewidth]{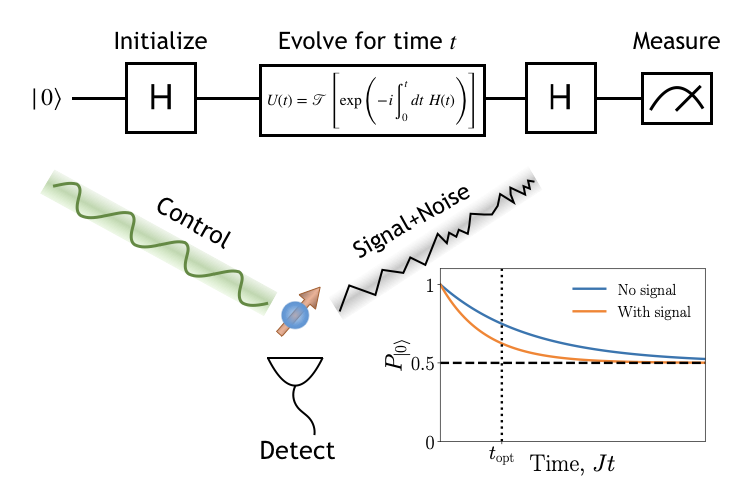}
\caption{Schematic description of the single-qubit experiment for detection of a signal. The qubit is initialized in an eigenstate of $\sigma_x$, $|\psi_{\rm in}\rangle=\frac{1}{\sqrt{2}}(|0\rangle+|1\rangle )$ and evolves in the presence of a signal, background noise and control. The measurement is done in the $\sigma_z$ basis after a Hadamard rotation. The average probability of the occupation of state $|0\rangle$ (denoted by $P_{|0\rangle}$) decays as a function of time, with the decay rate dependent on  the presence/absence of a signal. The detection is optimally performed at $t_{\rm opt}$ when the difference between the two decay rates is the largest.}
\label{fig:Schematic}
\end{figure}

While sensitivity is important for parameter estimation, it is less relevant for signal detection. In this manuscript, we reformulate the quantum sensing problem in a manner consistent with the following decision theoretic question: How does one optimally detect the presence of a stochastic signal with a known spectrum in the presence of background noise? The detection of signals in the presence of noise has been extensively studied in the field of classical decision theory \cite{10.2307/2235609}. A highly relevant applied formulation determined how to optimally choose whether a time-varying signal was signal plus noise or noise only, which has broad applicability for detection systems such as radar receivers~\cite{1954Peterson,Marcum1960}. This field of study was extended to the quantum domain by Helstrom \cite{HELSTROM1967254, Helstrom1969} who considered how to optimally choose between one of two density operators as the correct description of a receiver and Holevo \cite{HOLEVO1973337} who considered the question of the optimal measurement to distinguish between one of two quantum states. Results following these works have placed bounds on the limits of quantum state ~\cite{Audenaert2007,Calsamiglia2008} and channel discrimination~\cite{Acin2001,Sacchi2005,Wang2006,Hayashi2009,Harrow2010}. 

We build upon this early work but incorporate the filter function formalism originating from quantum control theory \cite{PhysRevLett.87.270405, PhysRevB.77.174509, PhysRevA.84.062323, Paz-SilvaPRL2014} to answer this question of how to optimally detect signals with a controllable quantum sensor. Unlike earlier works on state and channel discrimination that aim to identify the optimal measurements to distinguish two states/channels, our work is focused on identifying the control scheme that optimally separates the two cases.

Our model, as shown schematically in Fig. \ref{fig:Schematic}, considers a single qubit (or an ensemble of non-interacting qubits) with external fields coupled through an energy splitting term and single-axis control on the transverse axis. We seek control protocols that maximize the relative dephasing of the qubit under the two different scenarios of signal plus noise or noise only. We examine arbitrary single-axis time-dependent controls for this discrimination problem and show that under certain assumptions on the background noise, the optimal control protocol is remarkably just a constant control with Rabi frequency corresponding to the maximum of the signal spectrum. This corresponds to what is traditionally known as spin-locking in the nuclear magnetic resonance literature \cite{levitt2001spin}. The spin-locking technique, along with its pulsed analogue, has been widely used in quantum sensing and noise spectroscopy applications \cite{PhysRevA.86.062320, Aiello2013, PhysRevLett.110.017602, PhysRevLett.93.157005, PhysRevLett.109.153601, PhysRevB.85.174521}. 

{\it Model.}--- Let us consider a single qubit as a quantum sensor, in the presence of a dephasing signal and noise. Assuming uniaxial control (along $\sigma_x$), and under the rotating wave approximation, the qubit Hamiltonian is given by,
\begin{equation}
H(t)=\frac{1}{2}J\left[\sqrt{\alpha} \ s(t)+\eta(t)\right]\hat{\sigma}_z + \frac{1}{2}\Omega(t) \hat{\sigma}_x, \label{eq:Ham}
\end{equation}
where $\Omega(t)$ is the Rabi frequency of an arbitrary time-dependent control,  the signal~$s(t)$ and background noise~$\eta(t)$  are both considered to be classical wide sense stationary Gaussian stochastic  processes and $\alpha$ denotes the ratio of the signal-to-noise power~(SNR) and $J^2$ is the total noise power. These stochastic processes have mean zero, \mbox{$\overline{s(t)}=\overline{\eta(t)}=0$} and two-point time correlations given by  \mbox{$\overline{\eta(t)\eta(t^\prime)}=g_\eta(t-t^\prime)$}, \mbox{$\overline{s(t)s(t^\prime)}=g_s(t-t^\prime)$} and \mbox{$\overline{\eta(t)s(t^\prime)}=0$} with the normalization $g_\eta(0)=g_s(0)=1$ and $\overline{(\cdots)}$ denoting averaging over noise realizations. Alternatively, the noise correlations may also be represented in the frequency domain, using the power spectrum, $S_{\eta}(\omega)=\int_{-\infty}^{\infty}d\tau g_{\eta}(\tau)e^{-i\omega\tau}$, with the normalization $\int_{-\infty}^\infty d\omega S_\eta(\omega)=2\pi$, and similarly for $S_s(\omega)$. 
In this manuscript, we consider two distinct types of noise for the background. In the simple case of white-noise, we can obtain analytical results for the optimal control.  In addition, we also consider the background noise spectrum to be a Lorentzian, corresponding to $g_\eta(t)=e^{-|t|/\sigma_t}$, where $\sigma_t$ denotes the correlation time-scale.
This noise spectrum is quite relevant to quantum sensing platforms such as Nitrogen-vacancy centers in diamond~\cite{Klauder-Anderson1962,Bar-Gill2012}. We examine the Lorentzian case numerically.

\begin{figure}
\includegraphics[width=\linewidth]{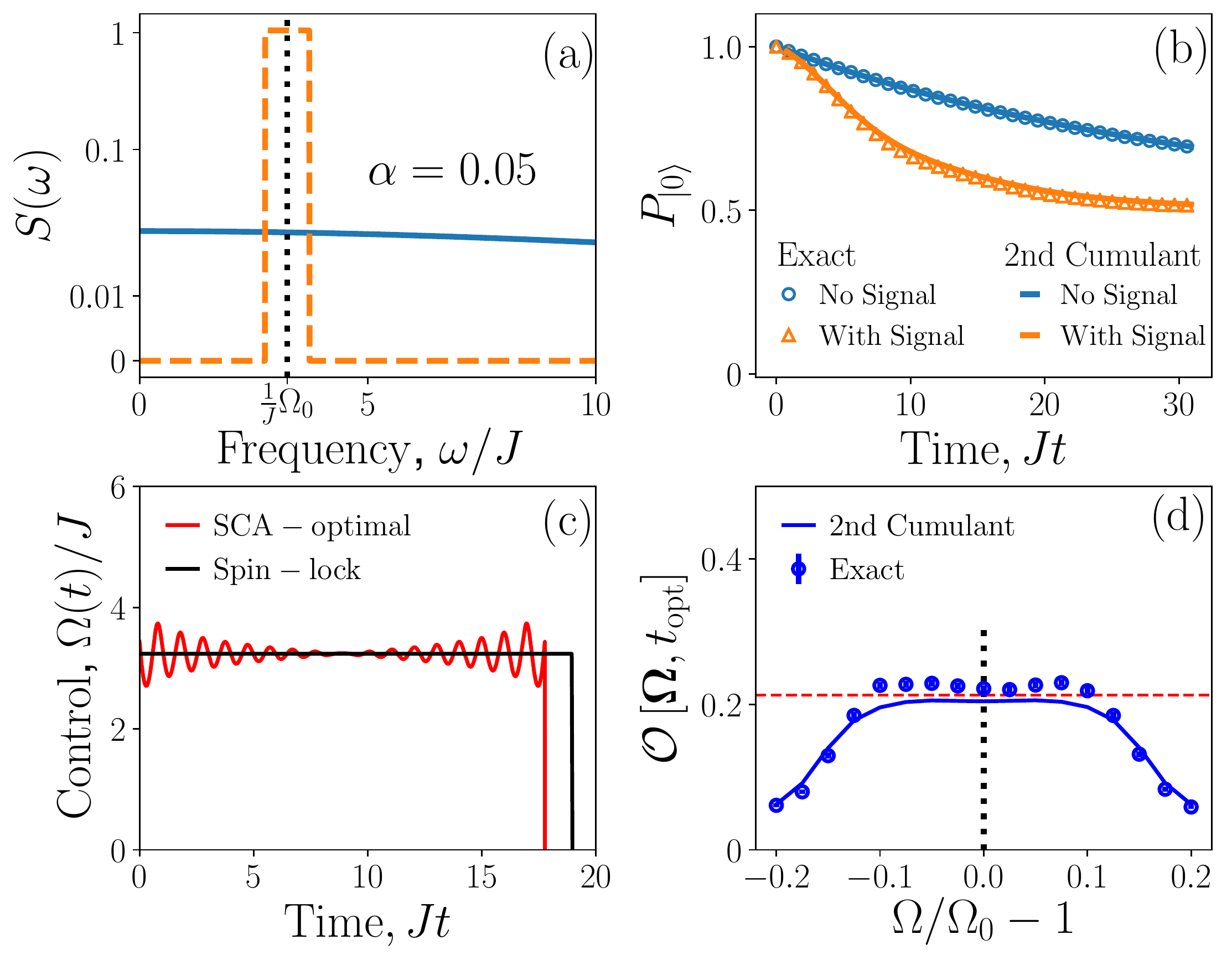}
\caption{Optimal signal detection for white background noise. (a) Background noise (blue) is considered to be nearly white ($J\sigma_t=0.01\sqrt{30/\pi}$),  and the signal spectrum (orange) is white cutoff centered around $\Omega_0/J=10\sqrt{\pi/30}$ and width $2\Delta \Omega/J= 6\sqrt{\pi/30}$ and SNR given by $\alpha=0.05$. (b) Dephasing under spin-locking, represented by outcome probability, $P_{|0\rangle}$, in the presence (orange) and absence (blue) of signal. The points and solid lines correspond to the outcome probabilities as calculated using exact numerical simulation and SCA respectively. (c) Comparison between  spin-lock (black) driving at frequency $\omega=\Omega_0$ to the numerically optimal control [See \cref{eq:optimal-Theta}] until their respective optimal time for measurement $t=t_{\rm opt}$. (d) Maximum of the objective function, $\mathcal{O}=\Delta P_{|0\rangle}$ [see \cref{eq:objective}] at $t=t_{\rm opt}$ for different spin-lock driving frequencies, $\Omega(t)=\Omega$, as obtained from exact numerics~(points) and from second cumulant expansion~(line). Red dashed line shows $\mathcal{O}_{\rm opt}=\mathcal{O}(\boldsymbol{\Omega}_{\rm opt},t_{\rm opt})$ for the SCA-optimal control [plotted in red in (c)].}
\label{fig:spin-lock}
\end{figure}

For the numerical simulations, the dynamics of the qubit sensor are modeled by an exact noisy quantum simulation and control library --`Mezze' \cite{mezze}, which 
uses a stochastic Liouville equation formalism with a Trotter decomposition.  We fix the noise power $J^2=30/\pi\approx 9.549$ and rescale all frequencies and time-scales in units of $J$. We also keep fixed the  SNR $\alpha=0.05$ and Trotter time-step $\Delta t=10^{-3}$.  The signal power-spectrum $S_s(\omega)$ is chosen to be white-cutoff, centered around a frequency $\Omega_0$ and width $2\Delta\Omega$; see e.g., \ref{fig:spin-lock}(a) (in orange). Note that the parameters considered in this manuscript are chosen to be in the regime of low SNR ($\alpha\ll 1$) and weak noise relative to the control ($J/{\rm max}\{\Omega(t)\} \lesssim 1$).  The weak noise limit allows us to utilize the cumulant expansion to study the dynamics of the qubit.

{\it Detection Protocol.}--- The aim of this protocol is to optimally detect the presence of a stochastic signal with a known spectrum in the presence of background noise. This is in contrast to the goal of estimating an AC or DC signal as traditionally considered in quantum sensing~\cite{RevModPhys.89.035002}. The detection protocol can be described in four steps. (i)~Initialize the qubit in the state $|+\rangle=\frac{1}{\sqrt{2}}(|0\rangle+|1\rangle)$. (ii)~Let the qubit evolve for time $t=t_{\rm opt}$ in the presence of the Hamiltonian, $H(t)$; see \cref{eq:Ham}. Here, $t_{\rm opt}$ is chosen depending on the signal, noise, and control by maximizing the infidelity between the signal present and signal absent cases. This maximizes the likelihood-ratio that allows one to optimally discriminate between the signal absent versus the signal present hypotheses by the Neyman-Pearson Lemma \cite{neyman1933ix}. (iii)~Rotate the qubit using a Hadamard gate (denoted as $\mathsf{H}$). (iv)~Measure in the $\hat{\sigma}_z$ basis. Record outcome as `$0$' or `$1$'. Repeat these steps $N_{\rm shots}$ times. This procedure is schematically described in Fig.~\ref{fig:Schematic}.  

In this paper, we make no assumptions on the shape of the control function $\Omega(t)$ in order to obtain the optimal detection protocol. The numerically optimized detection scheme, $\Omega(t)=\Omega_{\rm opt}(t)$ is compared with the performance of some well-known protocols used in sensing (i) Spin-locking: $\Omega(t)={\rm constant}$; (ii) Carr, Purcell, Meiboom, Gill (CPMG) pulse sequence: $\Omega(t)$ is given by a series of equidistant $\pi$-pulses separated by free evolution periods of duration $\tau_{\rm CPMG}$; and (iii) Ramsey interferometry: $\Omega(t)=0$. Examining the dephasing dynamics in the presence of control, the probability to observe the qubit in state $|0\rangle$ is $P_{|0\rangle}(t)=\frac{1}{2}(1+e^{-\chi(t)})$. Clearly, the dephasing exponent depends on the presence or absence of a signal and is denoted as $\chi_{\eta+s}$ and $\chi_{\eta}$ respectively. The corresponding outcome probabilities are labeled as $P_{\eta+s}$ and $P_\eta$ respectively.  In the following, we compute these two exponents in the regime of weak noise and signal as well as low SNR, and identify the control protocol that at a particular optimal time ($t_{\rm opt}$) of measurement gives the maximum separation between the two decaying average probabilities, $\Delta P_{|0\rangle}= P_\eta - P_{\eta+s}$.

{\it Dephasing in Second Cumulant Approximation (SCA).}--- Let us consider the dephasing of a qubit just in the presence of noise, setting $s(t)=0$. The dynamics of the qubit in a weak noisy environment, is well understood using the cumulant expansion~\cite{Paz-SilvaPRL2014,2017pazsilva:mq-qns}. The dephasing exponent, $\chi_\eta$, can be obtained from the dynamics  of $\overline{\langle\hat{\sigma_x}(t)\rangle}=\Tr\left[\sigma_x\rho(t)\right]\approx e^{-\chi(t)}\langle\sigma_x\rangle_0$; $\langle\sigma_x\rangle_0$ denoting the initial state. In the interaction frame of the control $U_I(t)=e^{-i\frac{1}{2}\hat{\sigma}_x\Lambda_t}$, where $\Lambda_t=\int_0^t d\tilde{t} \Omega(\tilde{t})$, the dynamics is straightforwardly rewritten in terms of the cumulant expansion, $\overline{\langle\hat{\sigma_x}(t)\rangle} =\Tr\left[\exp\left(\sum_{n=1}^{\infty}\frac{\left(-i\right)^{n}}{n!}C^{(n)}(t)\right)\rho(0)\sigma_x\right]$; $C^{(n)}$ denoting the $n$th cumulant \cite{2017pazsilva:mq-qns, supp}. In the regime that the noise is weak, it is sufficient to terminate the series at the second cumulant (odd cumulants vanish trivially) to obtain the following expression for the decay~\cite{supp},
\begin{align}
\chi_\eta(t)&\approx \frac{J^2}{2}\int_{0}^{t}dt_{1}\int_{0}^{t}dt_{2}\ e^{i\Lambda_{t_{1}}}g_\eta\left(t_{1}-t_{2}\right)e^{-i\Lambda_{t_{2}}}, \label{eq:chi-t}
\end{align}
where, $\chi_\eta (t)$ is always real. In the frequency domain, \cref{eq:chi-t} can be recast into an overlap between the noise spectrum and the Filter Function (FF), $\chi_\eta(t)=\frac{J^2}{2}\int_{-\infty}^{\infty}\frac{d\omega}{2\pi} S_\eta\left(\omega\right)|F_t\left(\omega\right)|^2$. 
The FF is now given by the following expression, 
 \begin{align}
 F_t\left(\omega\right)&=\int_{0}^{t}d\tilde{t}\ e^{-i\omega \tilde{t}+i\Lambda_{\tilde{t}}}, \label{eq:FF}
\end{align} 

where we use the symmetry of the noise spectrum, $S_\eta(\omega)=S_\eta(-\omega)$.  Note that the FFs have a normalization, $\int_{-\infty}^{\infty}\frac{d\omega}{2\pi}|F_t(\omega)|^2 d\omega=t$.

As an example, consider the limiting case when the background noise is white. In this case,  $g_\eta(t-t^\prime)\propto \delta(t-t^\prime)$, which is the Dirac $\delta$-function. To be consistent with our normalization [$g(0)=1$], we  define $g(t-t^\prime)=\lim_{\epsilon\rightarrow 0}e^{-|t|/\epsilon}$. However, to have a non-zero decay rate, we must rescale the noise power $J^2$ such that , $2J^2\epsilon=\gamma$ is a constant.  Now, $\chi_\eta$ is obtained straightforwardly from \cref{eq:chi-t}, $\chi_\eta(t)=\frac{1}{2}\gamma t$. Interestingly, this dephasing rate under white noise is constant regardless of the control, $\Omega(t)$ applied. This gives the standard $T_2$ time for the qubit with $T_2=2/\gamma$.

In order to compare with numerics, it is convenient to switch to the discrete time picture, with time-steps $\Delta t=t/N$, where $N$ is the total number of steps, in addition to a piecewise constant control $\boldsymbol{\Omega}=(\Omega_0,\cdots, \Omega_N)$. Here, \cref{eq:chi-t} becomes a Riemann sum and $\chi_\eta$ is a matrix expectation value. To this end, let us introduce (i) an $N-$dimensional vector for the control, $\boldsymbol{\Theta}_t=\frac{1}{\sqrt{N}}\left[e^{-i\Lambda_0},e^{-i\Lambda_{\Delta t}}, \cdots ,e^{-i\Lambda_{N\Delta t}}\right]^T$ with a normalization, $\boldsymbol{\Theta}_t^\dagger\cdot\boldsymbol{\Theta}_t=1$; and (ii) An $N\times N$ dimensional symmetric Toeplitz matrix, for the noise correlation function,  $\left[\mathbb{G}_{\eta}\right]_{ij}=g_\eta\left(\left(i-j\right)\Delta t\right)=\left[\mathbb{G}_{\eta}\right]_{ji}$. Using these definitions, the expression for $\chi$ can now be written in a compact form,
\begin{align}
\chi_{\eta}(t)&=\frac{1}{2}J^2t\Delta t\   \boldsymbol{\Theta}_t^{\dagger}\cdot\mathbb{G}_{\eta}\cdot\boldsymbol{\Theta}_t. \label{eq:chi-discrete}
\end{align}
The white-noise limit is recovered by setting $\mathbb{G}_\eta$ to be a diagonal matrix,$\mathbb{G}_\eta[i,i]=1$, and $J^2\Delta t=\gamma$. We also note that in this discrete time notation, the FF [See \cref{eq:FF}] is the Fourier series of $\boldsymbol{\Theta}$.

 {\it Optimization for Detection.}---Let us now describe the procedure for optimizing the control for detecting the signal. Both the signal and the noise cause the qubit to dephase, and the dephasing exponent is  straightforwardly obtained in the SCA using \cref{eq:chi-discrete}. Since the signal and  noise are uncorrelated, the decay in the presence of a signal is a sum, $\chi_{\eta+s}=\chi_\eta+\chi_s$, where  $\chi_s$ and $\chi_\eta$ are the decay exponents obtained from having just the signal or the noise present, respectively. Thus, the qubit decays at a faster rate  in the presence of both signal and noise compared to just the background noise. In order to optimize for detection, we maximize over the difference between the two outcome probabilities, $\Delta P_{|0\rangle}$.  We now have an effective heuristic for designing optimal detection controls that becomes optimal when the SCA applies. We define the following objective function,
 \begin{align}
\mathcal{O}(t,\{\boldsymbol{\Omega}\})&=\Delta P_{|0\rangle}=\frac{1}{2}e^{-\chi_\eta(t)}\left(1-e^{-\chi_s(t)}\right). \label{eq:objective}
 \end{align}
 We carry out the optimization as a two-step procedure: 
  (i) Optimize over  control trajectories to obtain $\boldsymbol{\Omega}_{\rm opt}=\rm{argmax}_{\{\boldsymbol{\Omega}\}}\left[\log\mathcal{O}\right]$ at a fixed detection time $t$. Recall that  the detection time sets the dimension of the control vector, ${\rm dim}\left[\boldsymbol{\Omega}\right]=t/\Delta t$. In the limit of white background noise, we will obtain analytically the optimal control that maximizes $\mathcal{O}$. More generally, for arbitrary noise spectrum, the optimization is implemented using stochastic gradient descent (SGD) algorithms. We do this using the Adam optimization algorithm  implemented in TensorFlow library~\cite{kingma2014adam,abadi2016tensorflow} taking advantage of graphical processing units to accelerate the optimization procedure. Note that it is also straightforward to add additional constraints on the variables that could be motivated by experimental requirements; e.g. maximum available power~\cite{supp}.
  (ii) Optimize over the time of detection, $t_{\rm opt}$ to obtain the optimal detection protocol. The optimal time is obtained using a grid search over different detection times $t$. Therefore, we obtain the optimal detector in the SCA, which we denote as `SCA-optimal', with $\mathcal{O}_{\rm opt}=\mathcal{O}(t_{\rm opt},\boldsymbol{\Omega}_{\rm opt})$

{\it White Background Noise}--- Let us start by discussing the SCA-optimal protocol when the background noise is close to white, $\sigma_t \ll 1/\Omega_0,1/\Delta\Omega$. Here, we  assume that the signal spectrum $S_s(\omega)$ is characterized by a peak at $\omega=\Omega_0$ and width $\approx 2\Delta\Omega$. This case allows us to compute the optimal control analytically. Dephasing in the absence of any signal, $\chi_\eta$, is independent of the applied control, which simplifies the optimization objective, $\mathcal{O}$~[See \cref{eq:objective}]. Therefore, $\mathcal{O}$  can be maximized by maximizing $\chi_s$ given by a formula analogous to \cref{eq:chi-t,eq:chi-discrete}.  Examining the expression of $\chi_s$ in terms of FFs, the following bound is obtained,
\begin{align}
\chi_s(t)\leq \frac{J^2\alpha t}{2}{\rm max}\left[S_s(\omega)\right],\label{eq:chi-bound}
\end{align}
where, we use the normalization, $\int_{-\infty}^\infty\frac{d\omega}{2\pi}|F_t(\omega)|^2=t$. In the following, we show that this upper bound on $\chi_s(t)$ can be achieved in the  limit of long times ($t\gg 1/\Delta\Omega$), using a control that is constant in time, $\Omega(t)={\rm argmax}_\omega\left[S(\omega) \right]=\Omega_0$.  This control scheme is commonly referred to as `{\it spin-locking}'.

A spin-locked (SL) control scheme, $\Omega(t)=\Omega_0$ has the FF, $F_t(\omega)=te^{i(\Omega-\omega)t/2}\text{sinc}\left[(\Omega_0-\omega)\frac{t}{2}\right]$  and decay exponent,
\begin{align}
\chi^{\rm SL}_s(t)=& \frac{J^2t^2\alpha}{2}\int_{-\infty}^{\infty} \frac{d\omega}{2\pi}\  S_s(\Omega_0+\omega) \text{sinc}^2\left(\omega t/2\right)\\
\stackrel{t\gg\frac{1}{\Delta\Omega}}{\approx}& \frac{J^2t^2\alpha}{2}\int_{-\pi/t}^{\pi/t} \frac{d\omega}{2\pi} S_s(\Omega_0+\omega)   \approx\frac{J^2\alpha t }{2}S_s(\Omega_0) \label{eq:chi-SL}
\end{align}
where, in the second step, we approximate the ${\rm sinc}(\epsilon x)$ as a rectangular function. Therefore, we see that the bound for $\chi_s(t)$[see \cref{eq:chi-bound}] can be saturated using this simple control protocol. While we have shown this result for a spectrum that has a maxima at $\Omega_0$, it is straightforward to generalize it to a white-cutoff signal spectrum [see Fig.~\ref{fig:spin-lock}(a)], in which case the optimal spin-locking frequency is at the middle of the band. We emphasize the following subtlety in the derivation of \cref{eq:chi-SL}:  The optimal time for measurement ($t_{\rm opt}$) is large compared to the inverse width of the signal spectrum, $t_{\rm opt}\gg1/\Delta\Omega$.

The SCA-optimal detection protocol at any particular time of measurement $t$, with white background noise, can also be calculated numerically without using time-consuming SGD based optimizers. Since this control maximizes $\chi_s$, it can be obtained from the eigenstructure of the correlation matrix $\mathbb{G}_s$ [see \cref{eq:chi-discrete}]. In fact, it is straightforward to show that $\chi^{\rm opt}_s(t)\leq \frac{1}{2}J^2\alpha t \Delta t \mathfrak{g}_{\rm max}$, where $ \mathfrak{g}_{\rm max}$~($\boldsymbol{\Phi}_t^{\rm max}$) denotes the largest eigenvalue~(eigenvector) of the correlation matrix of the signal spectrum $\mathbb{G}_s$. However, the eigenvector of the correlation matrix is not necessarily an allowed control protocol. Therefore, the SCA-optimal control protocol is obtained from identifying a control vector $\boldsymbol{\Theta}_t^{\rm opt}=\frac{1}{\sqrt{N}}\left[e^{-i\Lambda_0},\cdots , e^{-i\Lambda_t}\right]$ with the largest overlap with $\boldsymbol{\Phi}_t^{\rm max}$.  Noting, that the eigenspectrum of $\mathbb{G}_s$ is doubly degenerate (denote them by $i=\pm$), we can construct two possible SCA-optimal control vectors elementwise as,
\begin{align}
\left[\boldsymbol{\Theta}_{t,\pm}^{\rm opt}\right]_p=\frac{1}{\sqrt{N}}\frac{\left[\boldsymbol{\Phi}_{t,+}^{\rm max}\right]_p\pm i\left[\boldsymbol{\Phi}_{t,-}^{\rm max}\right]_p}{\left|\left[\boldsymbol{\Phi}_{t,+}^{\rm max}\right]_p\pm i\left[\boldsymbol{\Phi}_{t,-}^{\rm max}\right]_p\right|}.
\label{eq:optimal-Theta}
\end{align}
where, the two possible optimal controls correspond to driving around $\pm\Omega_0$. Having constructed the SCA-optimal control for arbitrary time $t$, the optimal time of measurement, $t_{\rm opt}$ is obtained by maximizing $\mathcal{O}$ from \cref{eq:objective}. Now, we have the SCA-optimal control for detection, $\boldsymbol{\Theta}_{t_{\rm opt}}^{\rm opt}$ an example of which is shown in \cref{fig:spin-lock}(c).

We numerically simulate the performance of the sensing protocol for white background noise; the results are shown in \cref{fig:spin-lock}. A nearly-white background  noise spectrum is obtained by choosing the correlation time to be small, $J\sigma_t=0.01\sqrt{30/\pi}\ll J/\Delta\Omega$. Fig.~\ref{fig:spin-lock}(b) shows the observed outcome probability of state $|0\rangle$, $P_{|0\rangle}$ as a function of time. Both with and without the signal, $P_{|0\rangle}$ (shown in orange and blue) decays exponentially, with  very good agreement between the exact dynamics (points) and SCA (solid line). We calculate the optimal time for detection $t_{\rm opt}$ by maximizing the difference, $\mathcal{O}$, see \cref{eq:objective}. In Fig.~\ref{fig:spin-lock}(c) and (d), we examine the optimality of spin-locking. Fig.~\ref{fig:spin-lock}(c) compares spin-locking to the numerically obtained SCA-optimal protocol, $\vec{\Theta}_{t_{\rm opt}}^{\rm opt}$ [see \cref{eq:optimal-Theta}], up to their respective optimal times of measurement. Comparing the two control schemes, it is clear that spin-locking is close to optimal. In fact, for long times of measurement, $t\rightarrow \infty$, $\boldsymbol{\Theta}^{\rm opt}_{t\rightarrow\infty}$ converges to spin-locking. However, at finite measurement times, $t=t_{\rm opt}$, $\boldsymbol{\Theta}^{\rm opt}_{t_{\rm opt}}$ remains  distinct from spin-locking, with oscillations around the spin-lock frequency. Fig.~\ref{fig:spin-lock} (d) compares the performance of spin-locking as a function of the Rabi frequency of the applied control $\Omega$ to the SCA-optimal detector. Comparing the performance of spin-locking as obtained from exact numerics (blue dots) with that obtained for the SCA-optimal control (red dashed line), it is clear that spin-locking may perform marginally better in practice, which we attribute to corrections to the dynamics beyond the SCA.

\begin{figure}
\includegraphics[width=\linewidth]{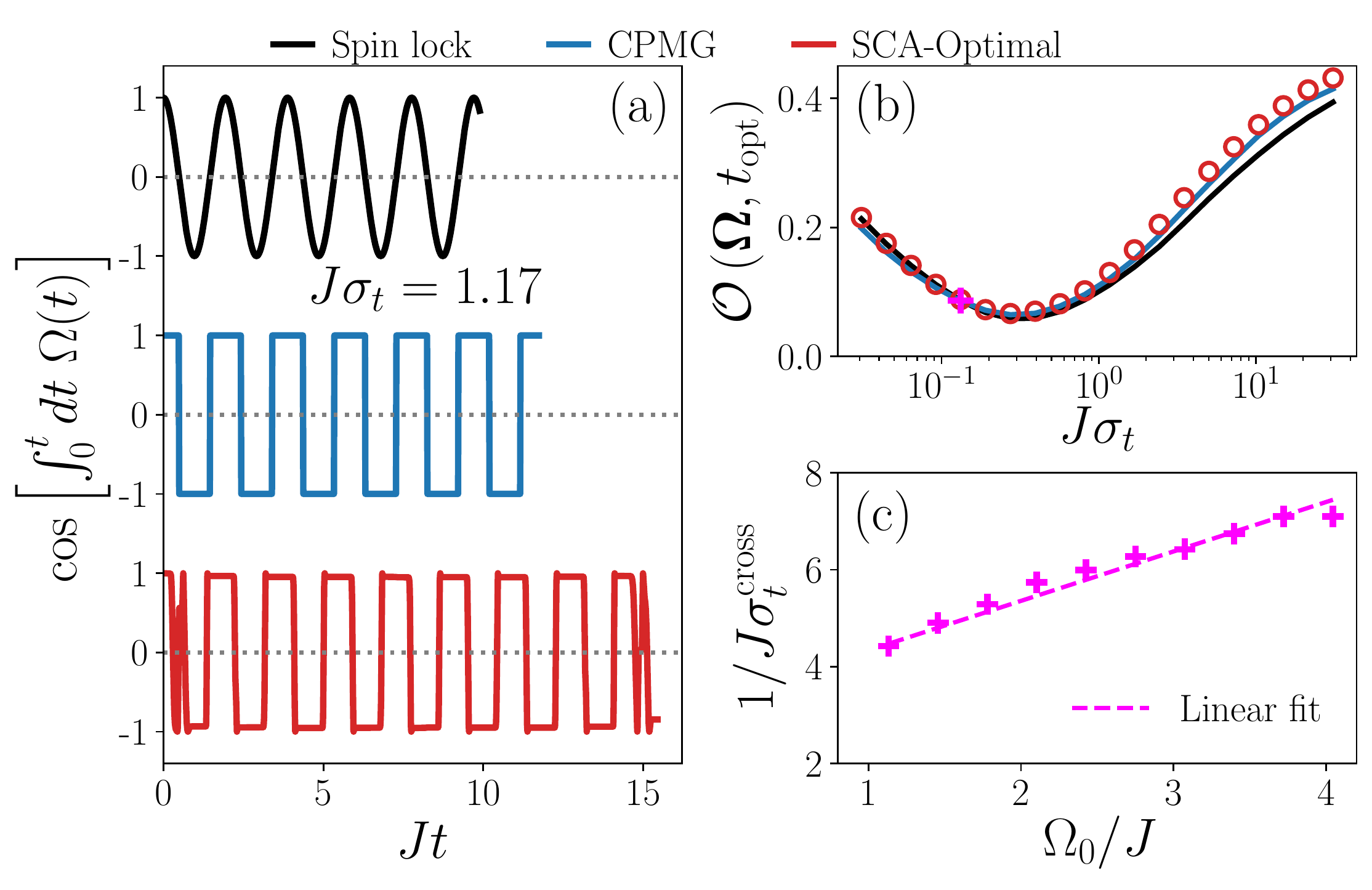}
\caption{Optimizing for detection for correlated background noise (Lorentzian spectrum) in the SCA. (a) Different control schemes shown until their corresponding optimal detection time, $t_{\rm opt}$: Spin-Locking (black), $\Omega(t)=\Omega_0$; CPMG (blue) with $\tau_{\rm CPMG}=\pi/\Omega_0$; and  Optimal control scheme (red) from the SGD based optimizer. The signal is at $\Omega_0/J=10\sqrt{\pi/30}$ and the correlation time of the background noise is $J\sigma_t=1.17$ (b) Maximum of the objective function $\mathcal{O}$ [see \cref{eq:objective}] as a function of background noise correlation, $\sigma_t$, for the three controls shown in (a). The signal is the same as in (a). The SCA-optimal control (shown as red circles) exhibits a crossover at $\sigma_t=\sigma_t^{\rm cross}$ (pink plus). It is close to Spin-Locking when $\sigma_t<\sigma_t^{\rm cross}$ and crossing over to CPMG when $\sigma_t>\sigma_t^{\rm cross}$. (c) The inverse of the crossover correlation time, $1/\sigma_t^{\rm cross}$ as a function of the center of the signal frequency $\Omega_0$ keeping the SNR $\alpha$ fixed. The dashed line is a linear fit to the points.}
\label{fig:comparison}
\end{figure}

\begin{figure}
\includegraphics[width=\linewidth]{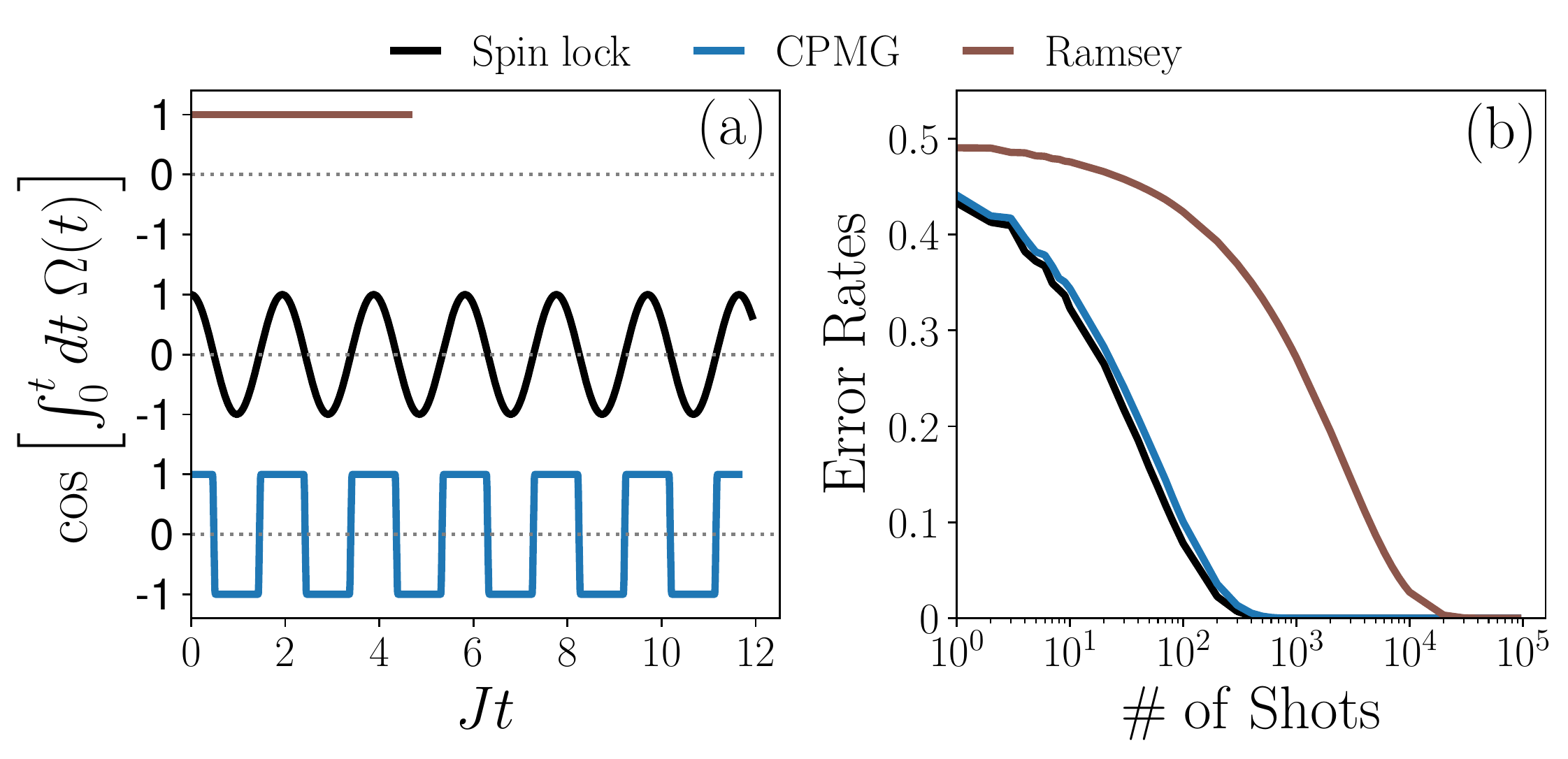}
\caption{(a) Comparison of different control schemes for a Lorentzian background spectrum using a full dynamical simulation of the qubit sensor. Signal is chosen centered at $\Omega_0/J=10\sqrt{\pi/30}$ and the background noise has the correlation time $J\sigma_t=1.17$. (b)  Error Rates for classification of the presence or absence of the signal given by the average probability of false positive and false negative outcomes,  as a function of the number of shots of the protocol (See \cref{fig:Schematic}). For a given number of shots, the threshold for the detector is chosen to minimize the total probability of FP and FN outcomes.}
\label{fig:comparison-numerics}
\end{figure}

{\it Lorentzian Background Noise.}--- We now compare the  performance of spin-locking as a function of correlation time of the background noise. Specifically, we compare it to CPMG with $\tau_{\rm CPMG}=\pi/\Omega_0$ and Ramsey interferometry, $\Omega(t)=0$. We also numerically obtain the SCA-optimal control protocol using SGD based optimizers and compare its shape with these protocols. The results are shown in \cref{fig:comparison,fig:comparison-numerics}. One of the important takeaways from the numerical optimization of the objective function,  $\mathcal{O}$, is that the SCA-optimal protocol depends on the correlation time, $\sigma_t$. The control obtained from the SGD, is close to spin-locking for short correlation times; however, for correlation times longer than a crossover scale, CPMG performs better~[See \cref{fig:comparison}(b)]. In \cref{fig:comparison}(a), we show the different controls for a particular correlation time, $J\sigma_t=1.17$. Clearly, the SCA-optimal control has close resemblance to CPMG. Interestingly, for larger correlation times, the SGD based optimizer finds novel control schemes that are neither CPMG or spin-locking~\cite{supp}.

The crossover correlation-timescale for  CPMG to perform better than spin-locking depends approximately as a power law to the signal frequency,  $\sigma_t^{\rm cross}\sim 1/\Omega_0$~[See \ref{fig:comparison}(c)]. The better performance of CPMG compared to spin-locking for longer correlation times can be understood qualitatively from the shape of their corresponding FFs. While the FF for spin-locking has all of its weight close to $\Omega_0$,  CPMG has some weight also at the odd harmonics of $\Omega_0$~\cite{alvarez2011:qns, Ajoy2011:cpmgff, supp}. In addition, the amplitude of the FF for CPMG at frequencies, $\omega<\Omega_0$, is lower compared to spin-locking~\cite{supp}. Since the background spectrum is Lorentzian and $S_\eta(\omega)$ decays at higher frequencies, this leads to $\chi_\eta^{\rm CPMG}<\chi_\eta^{\rm SL}$; unlike when the background is white where the $\chi_\eta$ is independent of the control. The optimization objective depends on $\chi_{s/\eta}$ in a non-linear fashion (see \cref{eq:objective}) which increases with either decreasing $\chi_\eta$ or increasing $\chi_s$.  Even though for the signal spectrum, $\chi_s^{\rm CPMG}<\chi_s^{\rm SL}$, the smaller value of $\chi^{\rm CPMG}_\eta$ leads to a larger $\mathcal{O}$.  It is interesting to note that, in the presence of experimentally motivated constraints, such as maximum available power, the optimal protocol interpolates between CPMG (instantaneous $\pi$-pulses) and spin-locking~\cite{supp}.

Let us now discuss the performance of the different control schemes as obtained from exact numerical simulation, see \cref{fig:comparison-numerics}. We compare the performance of spin-locking, CPMG and Ramsey schemes at their respective optimal detection times; the controls are shown in \cref{fig:comparison-numerics} (a). In order to quantify the performance of each control scheme, we calculate the rate of error for detecting the signal in the presence of background noise; see \cref{fig:comparison-numerics} (b). The error rate is defined as follows. Given $N_{\rm shots}$ measurement shots and a chosen threshold for detection (fixed number of $|0\rangle$ as measurement outcomes), the error rate of the detector corresponds to the average probability of False-Positive (FP) and False-Negative (FN) classifications. 
Choosing the threshold such that the average of probability of FP (type I error) and FN (type II error) is minimum, we plot the Error Rate as a function of the number of measurements in \cref{fig:comparison-numerics}(b); a better detector is characterized by lower error rates. Rather surprisingly, even when the correlation time is large $J\sigma_t=1.17$, the performance of CPMG and spin-locking remains fairly close to each other, with spin-locking still performing marginally better. This is likely due to the fact that the SGD based optimization is done under the assumption that the SCA holds. However, non-negligible contributions from higher cumulants to the dephasing rate leads to deviations from SCA, which may lead to better performance of spin-locking. This also reveals some general robustness in the optimality of spin-locking as a detection protocol for the signal.

{\it Discussion.}--- In this work we discuss the performance of different control schemes for detecting a known signal in the presence of certain background noise environments and show that a spin-lock drive is close to optimal in all cases we considered. This work opens up a potentially exciting use case for currently available quantum sensor hardware. These detectors can be used to identify signals in electromagnetic fields where the detection bandwidth is only limited by the frequency range of the control drive.

This work opens up several directions for future work. We only considered the dynamics in the second cumulant approximation where we are able to show the near-optimality of spin-locking. We did not take into account the role of the higher cumulants as they do not play a significant role at optimal detection time. However for qubits with higher $T_2$ times and larger signal power, it will be essential to consider its role for determining optimal controls for detection.  Another possibility is to consider the performance of detection protocols when the signal or noise is non-gaussian and/or non-stationary. Furthermore, we have considered the control drive to be on resonance. Any detuning will result in a two-dimensional control in the rotating frame, with controls along both $\sigma_x$ and $\sigma_y$. The optimal detection protocol in the presence of detuning will be a topic of future work, and will point to more robust protocols for detection. Finally, we have considered only a single qubit or a non-interacting ensemble of qubits as the sensor and do not consider the role of entanglement, which may provide an enhancement in sensing beyond that available classically.
\begin{acknowledgements}
 The authors would like to thank 
Mohammad Hafezi, Jeff Barnes, Colin Trout and Timothy Sweeney for useful discussions. PT, KMS, GDQ and BC also acknowledge funding from the Internal Research and Development program of the Applied Physics Laboratory. AS gratefully  acknowledges  support  from  ARO-MURI  and Physics Frontier Center by National Science Foundation at the JQI.
\end{acknowledgements}
\bibliographystyle{apsrev4-1}
\bibliography{QMF_refs} 



\section*{Supplementary material}

# Supplemental Material for
## "Optimal control for quantum detectors"

In this supplemental material we provide additional details for the main text. In Sec. S.I we derive the expression of the Filter Function, $F_t(\omega)$ defined in Eq. (3) in the main text. In Sec. S.II, we provide additional details for the numerical optimization of the objective function, $\Delta P_{|0\rangle}$, and dependence of the optimal control with the correlation time.

### S.I.  FILTER FUNCTION IN THE SECOND CUMULANT APPROXIMATION

In this section, we derive the expression for the decay exponent $\chi(t)$ and Filter Function, $F_t(\omega)$ defined in Eqs. (2) and (3) in the main text. For the purposes of this section, we set $s(t) = 0$. We follow the steps outlined in Refs. S1 and S2. Taking the definition of the Hamiltonian in Eq. (1) from the main text and transforming to an interaction frame with respect to the control, $U_c(t) = e^{-i\frac{1}{2}\sigma_x \Lambda(t)}$ we have,

$$H_I(t) = U_c^\dagger(t) H U_c(t) - i U_c^\dagger(t) i \partial_t U_c(t) \tag{S1}$$

$$= \frac{1}{2} J \eta(t) \left( \cos\left[\Lambda(t)\right] \sigma_z + \sin\left[\Lambda(t)\right] \sigma_y \right) \tag{S2}$$

where $\Lambda(t) = \int_0^t ds\, \Omega(s)$. Now, we define a new Hamiltonian, $\tilde{H}(s)$ [S1, S2],

$$\tilde{H}(s) = \begin{cases} -\sigma_x H_I(t-s)\sigma_x & 0 \leq s < t \\ H_I(t+s) & -t < s < 0 \end{cases} \tag{S3}$$

Now, we are ready to write down the time-evolution of particular observable. Of interest to us is the evolution of $\sigma_x(t)$ with a particular initial state $\rho(0)$. Noting that the observable $\sigma_x$ and the initial state $\rho(0) = |+\rangle\langle+|$, are unchanged in the interaction picture given by $U_c(t)$, we have under time evolution with $U_I(t) = \mathcal{T}\left[\exp\left(-i\int_0^t ds\, H_I(s)\right)\right]$,

$$\overline{\langle \sigma_x(t) \rangle} = \overline{\text{Tr}\left[\rho(t)\sigma_x\right]} = \text{Tr}\left[\overline{U_I \rho(0) U_I^\dagger}\, \sigma_x\right] \tag{S4}$$

$$= \text{Tr}\left[\overline{\sigma_x U_I^\dagger \sigma_x U_I}\, \rho(0)\sigma_x\right] \tag{S5}$$

$$= \text{Tr}\left[\overline{\mathcal{T}\left[e^{-i\int_{-t}^t \tilde{H}(s)\, ds}\right]}\, \rho(0)\sigma_x\right] \tag{S6}$$

where, we have averaged over noise realizations. Now, we can expand the exponential in the cumulant expansion,

$$\overline{\mathcal{T}\left[e^{-i\int_{-t}^t \tilde{H}(s)\, ds}\right]} = \exp\left(\sum_{n=1}^\infty \frac{(-i)^n}{n!} C^{(n)}(t)\right). \tag{S7}$$

The odd cumulants in the sum vanish since $H_I \propto \eta(t)$ and we assume that the noise is gaussian with vanishing mean.

*Second Cumulant Approximation*–When the noise is weak, it is sufficient to approximate the time-ordered exponential up to the second cumulant. We have the expression for the second cumulant,

$$C^{(2)}(t) = 2\int_{-t}^t dt_1 \int_{-t}^{t_1} dt_2\, \tilde{H}(t_1)\tilde{H}(t_2) \tag{S8}$$

$$= 2\int_0^t\int_0^{t_1} dt_1 dt_2\, \overline{H_I(t_1) H_I(t_2)} + 2\int_0^t\int_{t_1}^t dt_1 dt_2\, \sigma_x \overline{H_I(t_1)H_I(t_2)}\sigma_x - 2\int_0^t\int_0^t dt_1 dt_2\, \overline{\sigma_x H_I(t_1)\sigma_x H_I(t_2)}, \tag{S9}$$

$$\Rightarrow C^{(2)}(t) = 4\int_0^t\int_0^t dt_1 dt_2\, \overline{H_I(t_1) H_I(t_2)} \tag{S10}$$

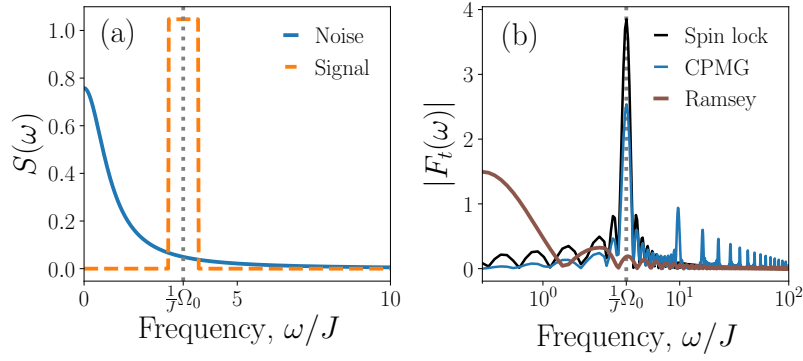

FIG. S1. (a) Noise and Signal spectrum corresponding to the parameters in Fig. 4 in the main text. (b) Filter Functions for the corresponding different control (Spin-Lock, CPMG and Ramsey) schemes; the time-domain profile of these controls are shown in Fig. 4(a).

The integrand can be simplified using,

$$H_I(t_1)H_I(t_2) = \frac{J^2}{4}\eta(t_1)\eta(t_2)\{\cos\left[\Lambda(t_1) - \Lambda(t_2)\right]\mathbb{I} + \sin\left[\Lambda(t_1) - \Lambda(t_2)\right]\sigma_x\} \tag{S11}$$

Averaging over noise realizations, and noting that $\overline{\eta(t_1)\eta(t_2)} = f_\eta(t_1 - t_2)$, with $f(t)$ being an even function. Therefore, the term $\propto \sigma_x$ in $C^{(2)}(t)$ vanish. Simplifying the expression for the second cumulant,

$$\Rightarrow C^{(2)}(t) = J^2 \int\limits_0^t \int\limits_0^t dt_1 dt_2 f_\eta(t_1 - t_2) \cos\left[\Lambda(t_1) - \Lambda(t_2)\right]\mathbb{I} \tag{S12}$$

Defining the decay exponent, $\chi(t)$ as $\overline{\langle\sigma_x(t)\rangle} = e^{-\chi(t)}\langle\sigma_x\rangle_0$, we have an expression for the decay from in the second cumulant expansion,

$$\chi(t) = \frac{J^2}{2}\int\limits_0^t\int\limits_0^t dt_1 dt_2 f_\eta(t_1 - t_2)\cos\left[\Lambda(t_1) - \Lambda(t_2)\right] \tag{S13}$$

$$= \frac{J^2}{2}\int\limits_{-\infty}^\infty \frac{d\omega}{2\pi}\, S_\eta(\omega)\frac{1}{2}\left(|F_{t,+}(\omega)|^2 + |F_{t,-}(\omega)|^2\right) \tag{S14}$$

$$F_{t,\pm}(\omega) = \int\limits_0^t d\tilde{t}\, e^{i\omega\tilde{t}\pm\Lambda(\tilde{t})} \tag{S15}$$

where we have used $S_\eta(\omega) = \int_{-\infty}^\infty d\tau f_\eta(\tau)e^{-i\omega\tau}$ and $f(\tau) = \frac{1}{2\pi}\int_{-\infty}^\infty d\omega S_\eta(\omega)e^{i\omega\tau}$. Note that, $F_{t,+}(\omega) = F_{t,-}(-\omega)$, which means that if $S(\omega) = S(-\omega)$, the expression for $\chi$ can be further simplified by introducing $F_t(\omega)$ introduced in Eq. (2). This completes the derivation for expressions of $\chi(t)$ and $F(\omega)$ as defined in Eq. (2) in the main text. As an example, we plot the filter functions to the controls considered in Fig. 4 (a).

## S.II. DETAILS FOR NUMERICAL OPTIMIZATION

We use a two-step numerical optimization to find the $\boldsymbol{\Omega}_{\text{opt}}$ that maximizes $\Delta P_{|0\rangle} = e^{-\chi_\eta}(1 - e^{-\chi_s})$. Specifically, we do a grid search over different durations of the protocol, and for each $t$, we find the control schedule that optimizes the objective function. Since the cost function is differentiable, we use the Adam [S3] optimizer with default parameters in TensorFlow to optimize the object function. Moreover, we constrain $[\boldsymbol{\Omega}]_i$ to be positive for all values of $i$. We either run the optimization for 10000 iterations, or stop if the magnitude of the difference of the objective function values separated by 1000 iterations is smaller than $10^{-6}$. In all the optimizations performed in this work, we discretize the control to $\Delta t = 0.01 J^{-1}$. We repeat this procedure for different control lengths $t$ from $300\Delta t$ to $1300\Delta t$ in increments of $10\Delta t$. We then choose a $t$ that has the optimal objective value.

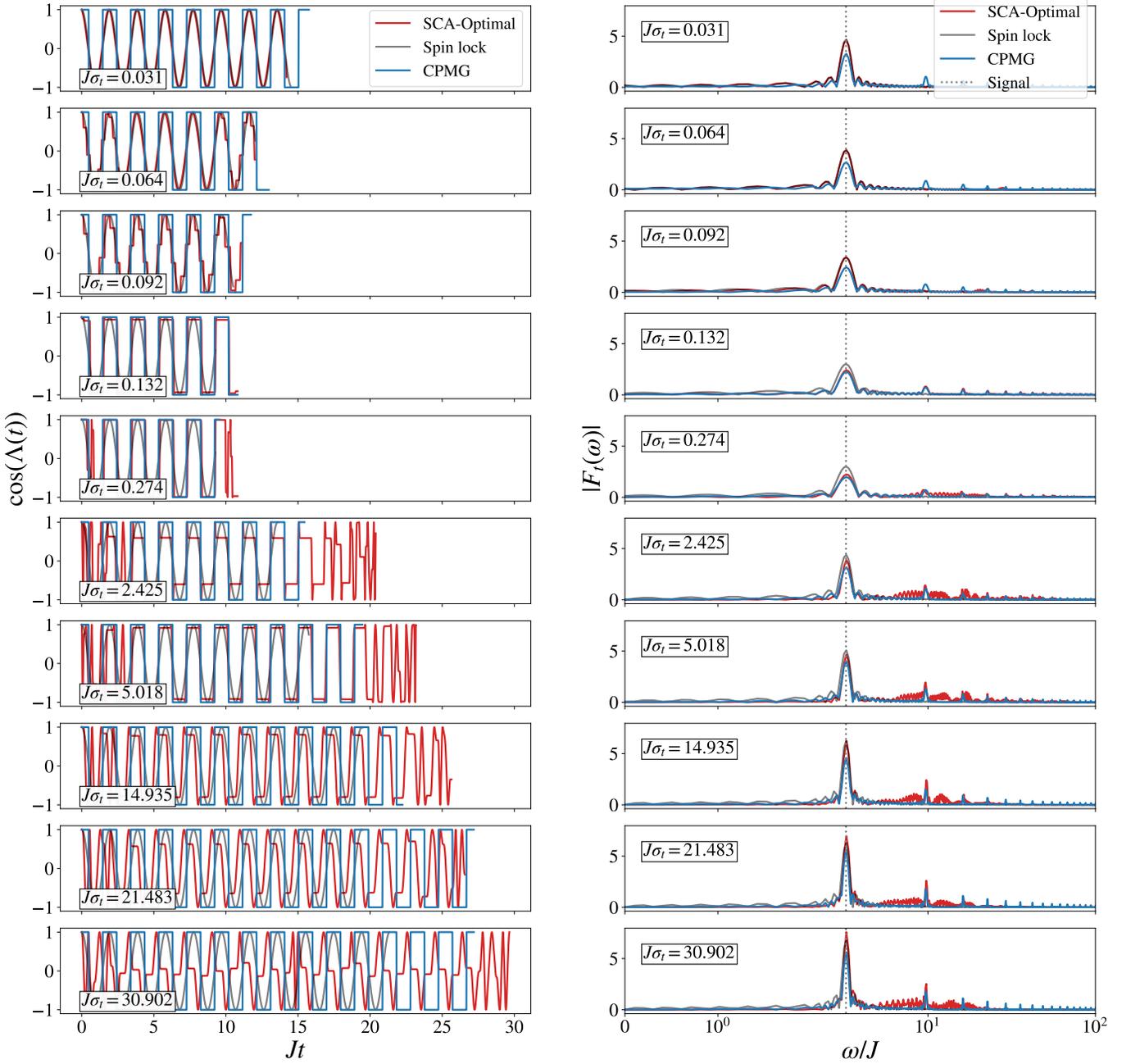

FIG. S2. Comparison of the numerically optimized protocol (SCA-optimal) with spin lock and CPMG protocols for different values of noise correlation $\sigma_t$. The left panel shows the time integrated control protocol $\cos(\Lambda(t))$, and the right panel shows the associated filter functions $|F_t(\omega)|$. Note that for small values of $J\sigma_t$, noise spectrum is nearly white and the optimal control coincides with spin-lock protocol. As $J\sigma_t$ increases the optimal control gets closer to CMPG protocol, and later assumes novel solutions that differ from both spin-lock and CPMG protocols.

After finishing this two-step optimization, we evaluate $\Delta P_{|0\rangle}$ at the optimal time with the optimal control by interpolating the results to a finer discretization of $\Delta t = 0.001J^{-1}$ and compare it to the value of $\Delta P_{|0\rangle}$ with CPMG and spin-lock control schemes. We obtain the best time for CPMG and spin-lock control schemes by numerically evaluating the $\Delta P_{|0\rangle}$ using discretized control with $\Delta t = 0.001J^{-1}$ and varying the duration of the protocol for $t$ from $3000\Delta t$ to $13000\Delta t$ in increments of $100\Delta t$. The optimal protocol compared with Spin-lock and CPMG for different values of $\sigma_t$, together with their corresponding filter functions, are shown in Fig. S2.

To take experimental limitations into account, we also consider a scenario where the maximum power is limited. This is manifested in a constraint on the maximum value of $\Omega$ at any time. We can implement this constraint by either adding an $L2$

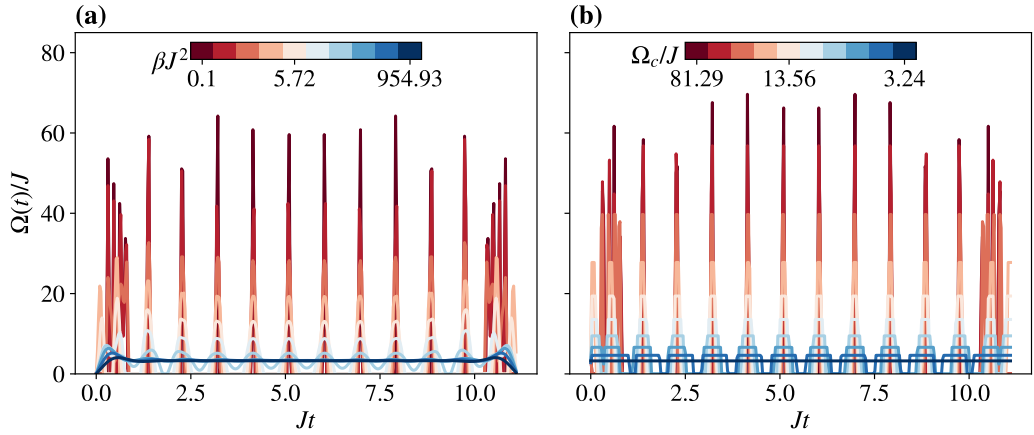

FIG. S3. The effect of constraints on the optimal protocol for $J\sigma_t = 0.56$. (a) With $L2$ regularization, we observe that as $\beta$ is increased, the pulse solution gets smoother and pulses are widened. (b) A hard cutoff $\Omega_c$ is imposed on the amplitude of the drive. As the cutoff value is decreased the instantaneous pulses are transformed to finite width pulses.

regularization term to the objective function, that is $\mathcal{O}_{L2} = \beta||\mathbf{\Omega}||_2^2$, where $\beta$ is a hyper parameter that controls the magnitude, or imposing a hard cutoff $\Omega_c$ on the value of the optimization variable. We observe that in a regime where the optimizer finds a CPMG-like control scheme to be optimal, adding these constraint transforms instantaneous pulses to finite-width version of the pulse that ultimately reaches the spin-lock solution, see Fig. S3.



\end{document}